\documentclass[a4paper]{article}

\usepackage[dvips]{graphicx}
\usepackage{amsmath}
\usepackage{journals_macros}
\usepackage{natbib}

\def\sna{SN\,Ia}
\def\apgt{\ {\raise-.5ex\hbox{$\buildrel>\over\sim$}}\ }
\def\aplt{\ {\raise-.5ex\hbox{$\buildrel<\over\sim$}}\ }
\newcommand{\ms}{\mbox {$M_{\odot}$}}
\newcommand{\mc}{\mbox {$M_{\mathrm{Ch}}$}}
\newcommand{\pyr}{\mbox {$\mathrm{yr^{-1}}$}}

\newcommand{\myr}{\mbox {~${\mathrm{M_{\odot}~yr^{-1}}}$}}
\newcommand{\md}{\mbox {$\dot{M}$}}
\newcommand{\nusn}{\mbox {$\nu_{\mathrm{Ia}}$}}
\newcommand{\ma}{\mbox {$\dot{M}_{\mathrm{a}}$}}

\bibpunct{(}{)}{;}{a}{}{,}

\begin{document}

\begin{center}
POPULATION SYNTHESIS FOR \\ PROGENITORS of TYPE IA SUPERNOVAE

Lev R. Yungelson

Institute of Astronomy of the Russian Academy of Sciences
\end{center}

%\email{lry@inasan.ru}

\begin{abstract}
We discuss application of population synthesis for binary
stars
to progenitors of SN\,Ia. We show that 
the only candidate systems able to support the
rate of SNe Ia
$\nusn \sim 10^{-3}$\,\pyr\ both in old  and young populations are merging white dwarfs.
In young populations ($\sim 1$\,Gyr) edge-lit detonations in semidetached systems
with nondegenerate helium star donors
are also able to support a similar \nusn.
The estimated current Galactic rate of SN\,Ia with single-degenerate progenitors is $\sim 10^{-4}$\,\pyr.
\end{abstract}

%------ body of article ------------------->>
\section{Introduction}

There is little doubt that explosions of SN\,Ia are thermonuclear disruptions
of the mass-accreting carbon-oxygen white dwarfs (CO WD) in binaries.  The main
facts arguing for this are: released energy per 1\,g is comparable to
$\epsilon_{CO \rightarrow Fe}$;  explosive nature of the events suggests that
degeneracy plays a significant role;  explosions may occur long after cessation
of star formation; hydrogen is not detected in the spectra of \sna\ (but see
discussion of single-degenerate scenario below). 

Identification of the  \sna\  progenitors is important for several reasons. It
may help to constrain the theory of binary--star evolution.   Modeling  and
understanding the explosions  requires knowledge of the initial conditions for
them and the environments in which they take place. Evolution of the galaxies
depends on the radiative, kinetic energy, and nucleosynthetic output of \sna\
and the evolution of \sna\ rate in time, which, in turn, depend on the nature
of the progenitor systems.  The nature of the progenitors is related  to the
use of \sna\ as distance indicators for determination of cosmological
parameters $H_0$ and $q_0$. Evolution of the luminosity function and the rate of SNe is important in this
respect.

A successful model for the population of progenitors of SN\,Ia has to explain
the inferred Galactic rate of events $(4\pm2)\cdot10^{-3}$\,\pyr\
\citep{captur01}, the origin of the observational diversity among local
($z<0.1$) SNe\,Ia --- $36\pm9$\% may be ``peculiar'' \citep{li+01}, and the
occurrence of SNe\,Ia  in stellar populations having a wide range of ages. 

Below, we discuss the scenarios of formation of binary systems in which \sna\
may occur and the rate of \sna, \nusn, predicted by different scenarios.

\section{Population synthesis}

The data provided by stellar evolution theory allows to construct  numerical
evolutionary scenario that describes the sequence of transformations of a binary
system with given initial masses of components and their separation $(M_{10},
M_{20}, a_0$) that it can experience in its lifetime.

Statistical studies of stars provide information on the binarity rate and the  distributions of binaries over $M_{10}$, $a_0$, $q_0=M_{20}/M_{10}$. Combined with star formation history, this allows to estimate the birthrate of  the systems with a given set of $(M_{10}, M_{20}, a_0$) at any epoch. Then,  
it is possible to compute
their contribution to the past or present population of stars of
different types. Integration over whole space of initial
parameters  or Monte Carlo simulation for a large sample of
initial ``binaries'' gives a complete model of the population of
binaries
and occurrence rates of different events, e. g., SN. 
Objects of the same type may be formed by several routes,
hence, one may expect variations 
of \sna.

\begin{figure}[!t]
\hspace*{1cm}
\includegraphics[scale=0.5]{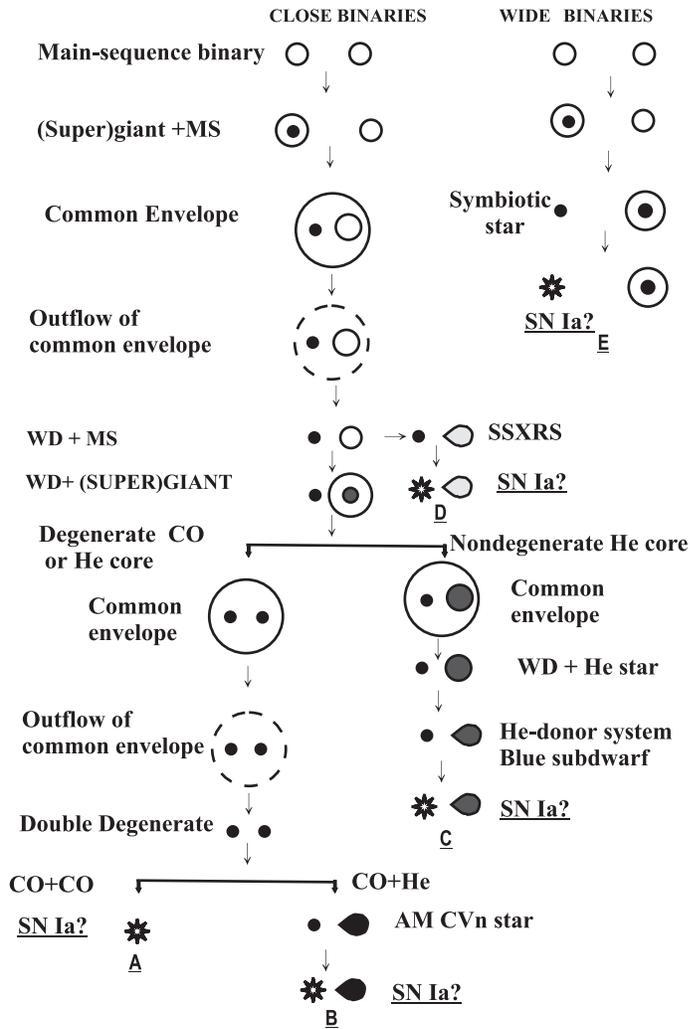}
\caption{Evolutionary scenarios for possible progenitors of SN\,Ia.}
\label{fig:scen}
\end{figure}

\section{Evolutionary scenarios for progenitors of SN\,Ia}

Figure \ref{fig:scen} shows (not to scale) a simplified flowchart of the main scenarios in which one may expect formation of a progenitor of \sna\ -- a CO WD that may ignite carbon in the center. The
 rates of formation of potential \sna\ via different channels are summarized in Table \ref{tab:occ}.

\begin{table}[!t]
\centering
\caption{Occurrence rates of SNe Ia in candidate progenitor systems (in \pyr)}
\hrule
\begin{tabular*}{\textwidth}{@{\extracolsep{\fill}}lcccccc}
Donor  & {CO WD}& {MS/SG} & {He star}  & {He WD} & {RG}  \\ \hline
 Counterpart & {Close} & {Supersoft} & {Blue} & {AM CVn} & {Symbiotic} \\
  &  {Binary WD} &  {XRS} & {sd} &  & {Star} \\ \hline
Mass transfer mode &{Merger} & {RLOF}     & {RLOF}  & {RLOF} & {Wind} \\ \hline
&\multicolumn{5}{c}{Direct carbon ignition (Chandrasekhar SN)}\\
{Young population}
& ${\bf 10^{-3}}$ & $10^{-4}$ & $10^{-4}$  & $10^{-5}$ &
$10^{-6}$  \\
{Old population}
& ${\bf 10^{-3}}$ & $ - $  & $ - $ & $10^{-5}$ & $10^{-6}$ \\
&\multicolumn{5}{c}{Edge-lit detonation (sub-Chandrasekhar SN)}\\
{Young population}
& $ - $ & $\aplt 10^{-4} $  & ${\bf 10^{-3}}$  &
$ - $ & $ \aplt 10^{-3} $ \\
\hline
\end{tabular*}
\label{tab:occ}
\end{table}

\textbf{Scenario A} [``double-degenerate''-- DD -- scenario, \citep{ty81,web84,it84a}] 
starts with a main-sequence (MS) binary
with $M_{10}, M_{20} \sim (4 - 10) $\,\ms.  The system is wide enough
for Roche-lobe overflow (RLOF) to occur when the
primary is an AGB star with a degenerate CO core. After
RLOF,
a common envelope (CE) forms. If components do not merge inside CE, the
core of the 
primary becomes a CO WD.
After dispersal of CE, the system remains wide enough for the secondary to become a CO WD too.
The angular momentum loss (AML) via gravitational waves radiation (GWR) results in the RLOF by the lighter of two WD.
Mass loss proceeds on dynamical time scale and in several orbital revolutions Roche-lobe filling WD turns into a disk around the more massive WD
\citep{ty79a,bbc+90}. If the total mass of the system exceeds \mc,  accretion from the disk may result in accumulation of \mc\ by the ``core'' and \sna. 

\textbf{Scenario B} is realized in the systems with $M_{20} \aplt 2.5\,\ms$ and such a separation of components after formation of the first WD that the secondary fills its Roche lobe in the hydrogen-shell burning stage and becomes a helium WD. Like in scenario {\textbf A}, dwarfs are  brought into contact by the AML
via GWR. Unstable merger, most likely, results either in ignition of He at the interface of accretor and disk \citep{efy01}, formation of a CE  and loss of He-rich matter or in formation of an R CrB-type star \citep{web84,ity96}.
If a stable semidetached system (of an AM CVn-type) forms, accumulation of \mc\ by accretor becomes possible.

In scenario \textbf{Scenario C} [``edge-lit detonation'' -- ELD -- scenario, \citep{livne90}]
$2.5 \aplt M_{20}/\ms \aplt 5$ and the separation between components after the first CE phase is such that the secondary 
fills its
Roche lobe before  core He ignition
and becomes a low-mass [$\simeq (0.35-0.8)\,\ms$] compact He-star. 
Low-mass helium remnants of stars
have lifetime comparable to the MS-lifetime of their progenitors. This allows AML via GWR to bring He-stars to RLOF before exhaustion of He in the cores. If mass loss occurs stably,
 $\ma \simeq (2-3)\cdot 10^{-8}$\,\myr, almost independent of the mass of companion \citep{skh86,tf89}.
Under  such \ma\
a degenerate He-layer forms atop WD and detonates when its mass increases to
$\sim 0.1\,\ms$ \citep{lt91}.
Detonation of He produces an inward propagating pressure wave that      
leads to close-to-center detonation of C.  The total mass of configuration in this case may be sub-Chandrasekhar.

\textbf{Scenario D} [``single degenerate'' -- SD -- scenario, \citep{wi73}] occurs in the systems where low-mass MS (or close to MS) stars [$M_{20}
\aplt (2 - 3)$\,\ms] or (sub)giant ($M_{20}/M_{1} \aplt 0.8$) companions to 
WD stably overflow Roche lobes. 
Accreted hydrogen burns into helium and then into CO-mixture. This allows
to accumulate \mc.

\textbf{Scenario E}  is the only way to produce \sna\ in a wide system, via accumulation of a He layer for ELD or \mc\ by accretion of stellar wind matter in a symbiotic binary \citep{ty_symb76}.

\begin{figure}
\hspace*{-0.5cm}
\includegraphics[scale=0.5,angle=-90]{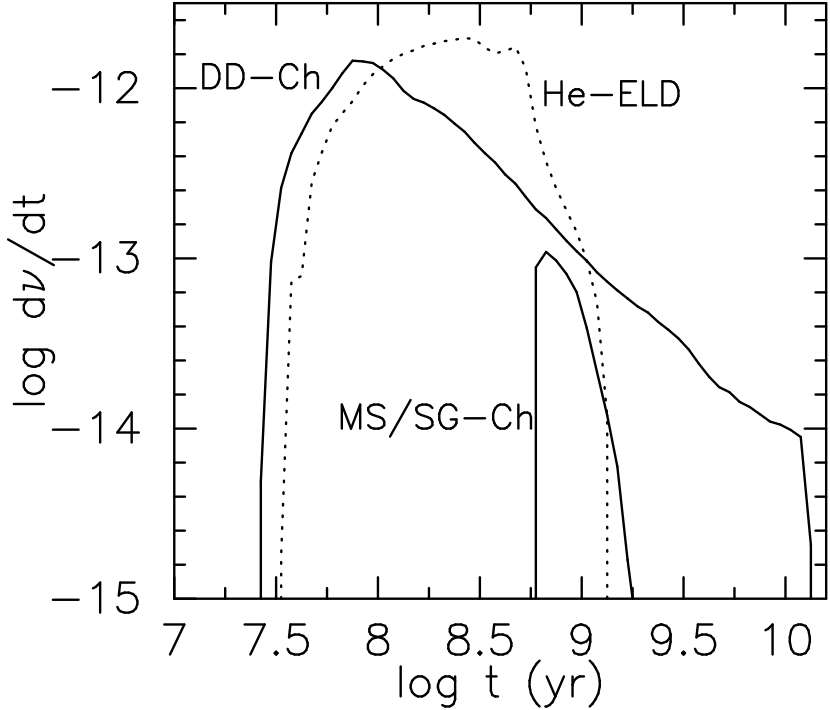}
\caption{Rates of potential \sna-scale events after a 1-yr long star formation burst that produces 1\,\ms\ of close binary stars.  
}
\label{fig:t_nu}
\end{figure}

Scenarios \textbf{A} -- \textbf{E} are associated with binaries of different types and with different masses of components. This sets an ``evolutionary clock'' -- the time delay between formation of a binary and  \sna. 
Figure~\ref{fig:t_nu} shows the differential rates of \sna\ produced via channels \textbf{A}, \textbf{C}, and \textbf{D} after a burst of star formation.
The DD-scenario is the only one that may operate in the populations of any age, while
SD- or ELD-scenarios are not effective if star formation ceased several Gyr ago.

Table \ref{tab:occ} presents the order of magnitude model estimates for \nusn\ after 10\,Gyr since beginning of star formation 
in the populations that have similar total mass comparable to the mass of the Galactic disk. Computations
were made by the code used, e. g., by Tutukov and Yungelson (1994) and Yungelson and Livio (1998) \nocite{ty94,yl98} for the value of common envelope parameter $\alpha_{ce}=1$.
(Differences in assumptions in 
population synthesis codes or parameters of computations result in numbers that vary by a factor of several; this is the reason for giving only order of magnitude estimates).
 ``Young'' population had constant star formation rate for 10\,Gyr;
in the ``old'' one the same amount of gas was converted into stars in 1\,Gyr.   We also list in the table the types of observed systems associated with certain channel and the mode of mass transfer.
Like Fig.~\ref{fig:t_nu}, table~\ref{tab:occ} shows that, say, for elliptical galaxies where star formation occurred in a burst, DD-scenario is
the only one 
able to respond for occurrence of \sna, while in giant disk galaxies with continuing star formation another 
scenarios
may contribute as well.

For a certain time the apparent absence of observed DD with $M_{\mathrm{tot}} \geq \mc$ merging in Hubble time was considered as the major ``observational'' difficulty for scenario \textbf{A}. Theoretical models 
predicted that it  may be necessary to investigate for binarity up to 1000 field WD with $V \aplt 16\div17$ for finding a  proper candidate
\citep{nyp+01}. The ``necessary'' number of WD was studied within SPY-project \citep{nap+01} and resulted in discovery of the first super-Chandrasekhar pair of dwarfs [R. Napiwotzki (this volume), Napiwotzki et al., (2003)\nocite{nap_spy03}].

On the ``theoretical'' side,
it was  shown for one-dimensional non-rotating models that the central C-ignition and \sna\ explosion are possible only for
$\md_{\mathrm a}\aplt(0.1 -0.2) \md_{{\mathrm{Edd}}}$ \citep{nomoto_iben85}.
But it was expected that in the
 merger products of binary dwarfs
 $\md_{\mathrm a}$ is close to $\md_{\mathrm{Edd}}\sim 10^{-5}$\,\myr\            \citep{mochko90} because of high viscosity in the transition layer between the core and the disk. For 
such  $\md_{\mathrm a}$ 
the nuclear burning will start at the core edge, propagate inward and convert the dwarf into an ONeMg one. The latter  will collapse 
without \sna\ \citep{isern+83}. However,
consideration of the role of deposition of angular momentum into central object (Piersanti et al., 2003a,b) \nocite{piers+03b,piers+03a}
has shown that, as a result of spin-up of rotation of WD, instabilities associated with rotation, deformation of WD and angular momentum loss by distorted configuration via GWR, $\md_{\mathrm a}$ 
that is initially $\sim 10^{-5}$\,\myr, decreases to
$\simeq 4\cdot1 0^{-7}$\,\myr. For this $\md_{\mathrm a}$ close-to-center ignition of carbon becomes possible.

\begin{figure}
\hspace*{0.5cm}
\includegraphics[scale=0.5,angle=-90]{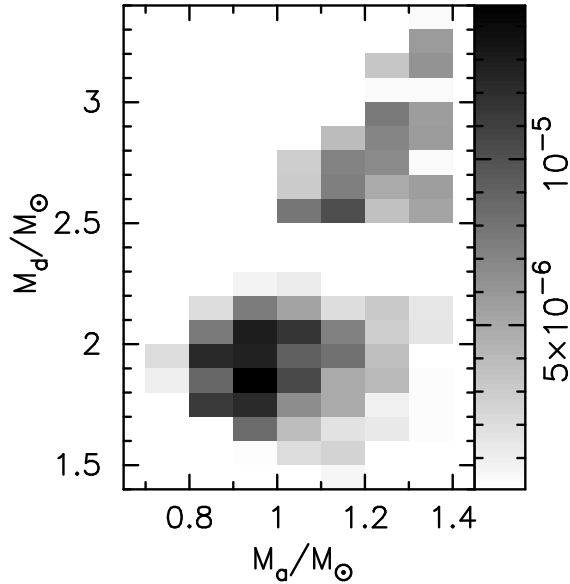}
\caption{The rate of accumulation of \mc\ in the SD-scenario (in \pyr), depending on 
the masses of WD-accretors and MS- or SG-donors at the beginning of accretion stage. }  
\label{fig:mamd}
\end{figure}

Because of long apparent absence of an observed  ``loaded gun'' for the DD-scenario  and 
its ``theoretical problems'',
SD-scenario (\textbf{D}) is often considered as the most promising one. However, it also encounters severe problems. 
No hydrogen is observed in the spectra of \sna,   while it is expected that
 $\sim 0.15$ \ms\ of H-rich matter may be stripped from the companion by the SN shell \citep{marietta00}\footnote{Recently discovered \sna\ 2001ic and similar 1997cy \citep{hamuy03} may belong to the so-called SN\,1.5 type or occur in a symbiotic system \citep{cy04}.}. Hydrogen may be discovered both in very early and late optical spectra of SN and in radio- and X-ray ranges \citep{eck+95,marietta00,lentz+02}. As well, no expected \citep{marietta00,canal+01,pods03} high luminosity and/or high velocity former companions to exploding WD were discovered as yet.

In the SD-scenario, hydrogen first burns into helium and then into C/O mixture. However, two circumstances hamper accumulation of \mc. \\
  At 
$\ma \aplt 10^{-8}$\,\myr all accumulated mass is lost in Novae explosions \citep{prkov95}. 
Even if \ma\ allows accumulation of He-layer, most of the latter is lost after He-flash 
\citep{it96symb,cassisi_etal98,piers+99},
dynamically or via frictional interaction of binary components with giant-size 
CE. Thus, the results of computations strongly depend on the assumptions on the amount of mass loss in the nuclear-burning flashes. The flashes become less violent and more effective accumulation of matter may occur if mass is transferred on the rate close to the thermal one \citep{it84a,yl98,ivanova_taam03}.
 However, this assumption seems to lead to overproduction of supersoft X-ray sources [see the estimate of the number of sources in \citet{fty04} and completeness of surveys estimates in 
\citet{stefrap95}]. 

The ``favorable'' range of mass transfer rates widens if mass exchange is stabilized by optically thick stellar wind from WD \citep{hkn96}. Under this assumption (not based on a rigorous treatment of the radiation transfer),  the excess of transferred matter over the upper limit for stable hydrogen burning ($\simeq 5\cdot10^{-7}$\,\myr\ for a 1\,\ms WD) is blown out of the system taking away specific angular momentum of the WD. This allows to avoid formation of CE for mass transfer rates up to $\simeq 10^{-4}$\,\myr\ and, simultaneously, implies stable hydrogen burning and reduces mass loss in helium burning flashes. 
Figure \ref{fig:mamd} shows the range of masses of donors and accretors in ``successive'' \sna\ progenitors at the beginning of accretion onto the WD stage, obtained under ``stabilization'' condition and for thermal-time scale mass transfer by Fedorova et al. (2004)\nocite{fty04}. The maximum of \nusn\ in the latter study 
is $2 \cdot 10^{-4}$\,\pyr, i. e., it still does not exceed $\sim10\%$ of the inferred Galactic \nusn.  Han \& Podsiadlowski (this volume) obtain for this channel the rate up to $1.1 \cdot 10^{-3}$\,\pyr, closer to the observational estimate.

\begin{figure}
\hspace*{-1.5em}
\includegraphics[scale=0.5,angle=-90]{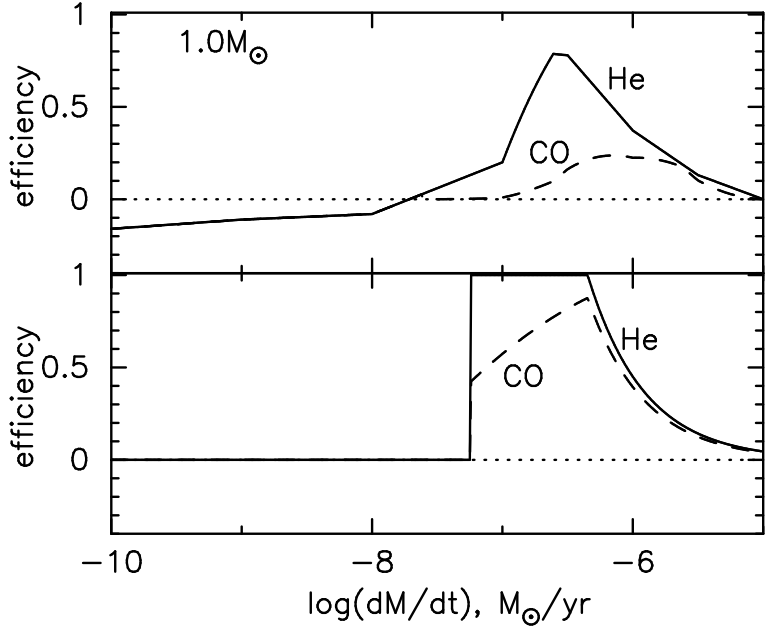}
\caption{Comparison of efficiency of accumulation of matter by a 1\,\ms\ white dwarf under different assumptions. See text for details.}  
\label{fig:compef}
\end{figure}

An important source for 
 discrepant \nusn\ obtained for SD-scenario may be the difference in the assumptions on the mass accumulation efficiency. As an extreme example, the upper panel of Fig. \ref{fig:compef} shows the efficiency of accumulation of He and C+O if one takes into account stellar wind mass loss by dwarfs that burn hydrogen steadily, mass loss in Novae explosions after Prialnik and Kovetz (1995) \nocite{prkov95} and estimates of mass loss in helium flashes after Iben and Tutukov (1996)\nocite{it96symb}; the lower panel shows efficiency of accumulation under prescriptions 
adapted by Han \& Podsiadlowski
[Fedorova et al. (2004)  implemented an ``intermediate'' case: assumptions on H-accumulation after Prialnik and Kovetz
and assumptions on He-burning similar to Han \& Podsiadlowski].

Scenario \textbf{C} may operate in populations where star formation have ceased no more than $\sim 1$\,Gyr ago and produce
SN at the rates that are comparable with the Galactic \nusn. But the outcome of ELD currently seems to be not compatible with observations of \sna. ``By construction'' of the model, the most rapidly moving products of explosions have to be He and Ni;
this is not observed.  The spectra produced by ELD are not compatible with observations of the overwhelming majority of \sna\ \citep{hk96}. On the theoretical side, it is possible that lifting effect of rotation that reduces effective gravity and degeneracy in the helium layer may prevent detonation  \citep{langer+03}.

Channel \textbf{B} most probably gives a very minor contribution to the total \sna\ rate since typical total masses of the systems are well below \mc.

The peculiarity of channel \textbf{E}
is the behavior of  $\md_{{\mathrm{a}}}$ from the wind: it is initially very low and
grows, as companion to the WD expands. Typical initial masses of WD in symbiotic stars are well below 1\,\ms\ \citep{yltk95}. For them it is more likely to accumulate a 
helium layer that may be lost in a thermal flash than  accumulate \mc. 

\section{Conclusion}

1. Only DD may secure the observed \nusn\ both in old and young populations. Merging pairs with $M_1+M_2 \simeq \mc$ were discovered after search in a WD-sample of appropriate size.
Account for effects of rotation may solve the problem of central ignition in the merger product.
Crucially needed is a study of the physics of merger which follows development of shocks and turbulence in the ``transition'' zone, transfer of momentum, rotation effects upon evolution of the ``core-disk'' configuration.

2. Edge-lit detonations in He-accreting systems can be responsible for \sna-scale events 
only in the populations younger than $\sim1$\,Gyr.
Lifting effect of rotation may reduce the number and scale of ELD.

3. Single-degenerate scenario may contribute a fraction ($\sim\,10\%$) of all events in young or intermediate age populations.
The major obstacle to SD-scenario are H and He thermal flashes.
Predictions of the rate of SD-events have to be reconciled with the number of supersoft X-ray sources.
A crucial test for SD-scenario would be  detection of H which may be present due to the interaction of SN shell with companion  or a ``slow wind'' of pre-SN. 

4. In the DD-scenario one may expect that exploding objects would differ in mass and central C-abundance. In SD-scenario all exploding WD most probably have \mc, but differ in central C. It is unclear whether these differences may explain  the diversity of observed \sna.

\vskip 0.3cm

This study was supported by RFBR grant 03-02-16254 and Federal Program ``Astronomy''.
The author acknowledges financial support
of IAU and JD5 SOC that enabled his participation in the meeting.

\nopagebreak

%\bibliography{journals,binaries}
%\bibliographystyle{apj}

\end{document}